# Seismology with optical links: enabling a global network for submarine earthquake monitoring

Giuseppe Marra[1], Cecilia Clivati[2], Luckett Richard[3], Anna Tampellini[2,5], Jochen Kronjäger[1], Louise Wright[1], Alberto Mura[2], Filippo Levi[2], Stephen Robinson[1], André Xuereb[4], Brian Baptie[3] and Davide Calonico[2]

**Earthquake monitoring across the globe is currently achieved with networks of seismic stations. The data from these networks have been instrumental in advancing our understanding of the Earth's interior structure and dynamic behaviour. However, almost all seismic stations are located on land and earthquakes of magnitude smaller than 4 at the bottom of the oceans remain largely undetected. Here we show that ordinary telecommunication optical fibre links can detect seismic events when combined with state-of-the-art frequency metrology techniques. We have detected earthquakes over terrestrial and submarine optical fibre links with length ranging from 75 to 535 km and a geographical distance from the earthquake's epicentre ranging from 25 to 18,500 km. In contrast to existing commercial reflectometry-based acoustic sensing methods used widely in the oil and gas industry[1–3], which are limited to only a few tens of kilometres, the technique presented here can be extended over thousands of kilometres, paving the way for detection of remote underwater earthquakes. By using the proposed technique on the existing extensive submarine optical fibre infrastructure, which already criss-crosses the seas and oceans, a global seismic network for real-time detection of underwater earthquakes could be implemented. The ability to detect off-shore earthquakes closer to the source could also enable a cost-effective solution for early detection of tsunamis.**

Underwater seismic sensors, such as Ocean Bottom Seismometers (OBS), have been widely used to better understand the physics of the Earth[4], from earthquake dynamics to changes in volcanic structure[5], magma generation and mid-ocean ridge development[6]. These devices are deployed over geographically limited areas for temporary surveys and data is recovered only once they are retrieved at the end of the measurement campaign[7–9]. A small number of countries, such as Japan, U.S.A. and Canada, have installed permanent arrays of OBSs close to earthquake-prone areas for both research purposes and as tsunami alert systems[10–13]. However, a permanent array of wired OBSs large enough to cover the Earth's waters, which account for 70% of the planet's surface, would be prohibitively expensive to implement. Its cost has been estimated between 700 million and 1 billion Dollars[14]. Thus, in the last decade, other more affordable solutions have been proposed and implemented which employ gliders[15] and free-floating buoys[16]. A large task force has also been setup to define a road-map to include sensors in future submarine telecommunication repeaters[17]. However, the existing vast submarine telecommunication network already provides the infrastructure to implement a global real-time seismic network if the fibre itself is used as the sensing element.

Submarine optical fibre cables are the backbone of international and intercontinental telecommunication. Since the first installations in the 1990s, the growth of the Internet and mobile services has caused the number of submarine fibre links to increase exponentially, to a current total length of over one million km. In 2016 alone, approximately 100,000 km of cable were added to the existing network and another 200,000 km are planned to be installed by mid-2018[18]. A map showing the current and planned submarine coverage is shown in Fig. 1.

[1]National Physical Laboratory, Hampton Road, Teddington, TW11 0LW, UK. [2]I.N.Ri.M., Istituto Nazionale di Ricerca Metrologica, Strada delle Cacce 91, 10135 Turin, Italy. [3]British Geological Survey, The Lyell Centre, Research Avenue South, Edinburgh, Scotland, UK, EH14 4AP. [4]Department of Physics, University of Malta, Msida MSD 2080, Malta. [5]Politecnico di Torino, Corso Duca degli Abruzzi 24, 10129, Turin, Italy

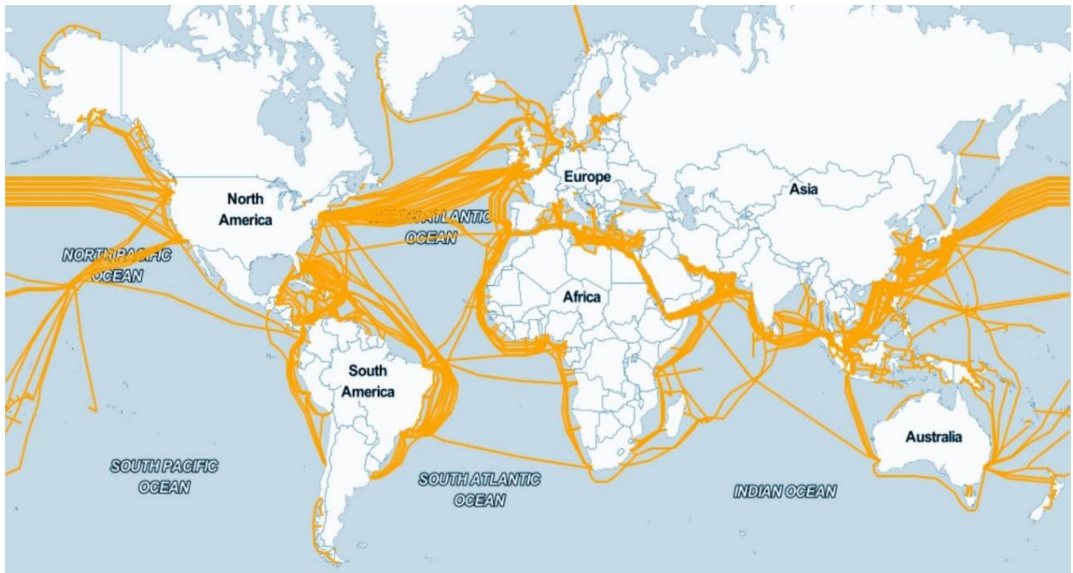

**Figure 1 | Submarine telecommunication cable map**. Illustration of the existing and planned submarine telecommunication infrastructure. Optical frequency metrology techniques enable these fibre links to be used for the detection of small earthquakes at the bottom of seas and oceans which otherwise are too weak to be detected by seismometers on land. Map data © OpenStreetMap contributors; Cable data: TeleGeography's Telecom Resources licenced under Creative Commons Share alike.

Whilst primarily designed for telecommunication, optical fibre links are now finding new applications. During the last decade, time and frequency metrology laboratories around the world have been developing advanced techniques for using optical fibre links to compare the frequencies of next-generation atomic clocks. In Europe, fibres up to 2,200 km long already link most of the largest National Measurement Institutes and expansion of the current network is underway[19–22]. It is now possible to measure changes as small as a few femtoseconds in the propagation delay experienced by the laser light travelling in the fibre, corresponding to length changes at the 1 μm-level, even over thousands of kilometres of fibre. This extraordinary level of sensitivity, which is achieved in just 1 s of measurement time, is enabled by frequency-metrology grade lasers which generate phase-stable light over the entire propagation time through the fibre. This ensures that any propagation delay change measured at the end of the link can be attributed exclusively to the fibre. The phase stability of these state-of-the-art lasers is sufficient to enable coherent measurements over fibre lengths well beyond 10,000 km. The propagation delay changes are caused by environmental perturbations to the fibre, such as vibrations, acoustic noise and temperature fluctuations. Fibre links are usually installed in underground utility ducts, such as power or gas lines, or along motorways and are thus exposed to all these perturbations. The induced noise is detrimental to atomic clock comparisons and is suppressed using active cancellation techniques. However, this unprecedented level of sensitivity to environmentally-induced perturbations can be exploited to detect seismic waves, and potentially any other source of vibrations and acoustic noise, not only on land but also at the bottom of the sea if submarine cables are used.

On 24$^{th}$ August 2016 an earthquake of momentum magnitude (Mw) 6.0 struck in Central Italy, followed by two more events of Mw 5.9 and Mw 6.5 on 26$^{th}$ and 30$^{th}$ October. These seismic events were detected at the National Physical Laboratory (NPL) in Teddington (UK) whilst running frequency metrology experiments on an optical fibre link not intentionally designed to detect seismic waves. This 79 km-long fibre link (UK-L1) connects the National Physical Laboratory in Teddington (UK) to a data centre in the nearby town of Reading and is located at a geographical distance of approximately 1,400 km from the epicentre of the Central Italy earthquake. The phase fluctuations induced by the seismic event on the laser light propagating in the fibre link for the 30th October event are shown in Fig. 2a and compared to data from a seismic station (Swindon, GB.SWN1) located approximately 100 km away from the NPL end of the fibre link. Whilst the magnitude of the detected primary wave (P-wave), the frequency spectrum of which extends to a few Hz, is partially suppressed by the low phase sampling rate (1 s) used at the time, the arrival time of the P-wave can still be determined. Subsequently, several other tele-seismic events were detected, with epicentres in New Zealand, Japan and Mexico and magnitude ranging from 5.9 to 7.9. A higher signal-to-noise ratio of the detected seismic waves was achieved on another 75 km-long optical link (UK-L2) in South-East England in late 2017. UK-L2 runs almost entirely in non-metropolitan areas, resulting in lower environmentally-induced noise levels compared to UK-L1. At the same time, a link (IT-L1) was also established by the Istituto Nazionale di Ricerca Metrologica (INRIM) in Turin (Italy), between Turin and Medicina near the town of Bologna, with a length of 535 km. In both UK-L2 and IT-L1 links the optical phase was sampled at a rate of 100 samples/s, enabling better timing resolution compared to previous measurements on UK-L1. Fig. 2b shows a Mw 7.3 earthquake on the Iraq-Iran border (12th November 2017) as detected on both UK-L2 and IT-L1 links, compared to signals detected by nearby seismometers. The arrival times for the P and S waves can be accurately determined using the UK-L2 link and are compatible with those identified using the nearby seismic station at Herstmonceux (GB.HMNX). In the IT-L1 link periodic environmental perturbations make it difficult to resolve the P-wave whilst the following seismic perturbations are clearly visible.

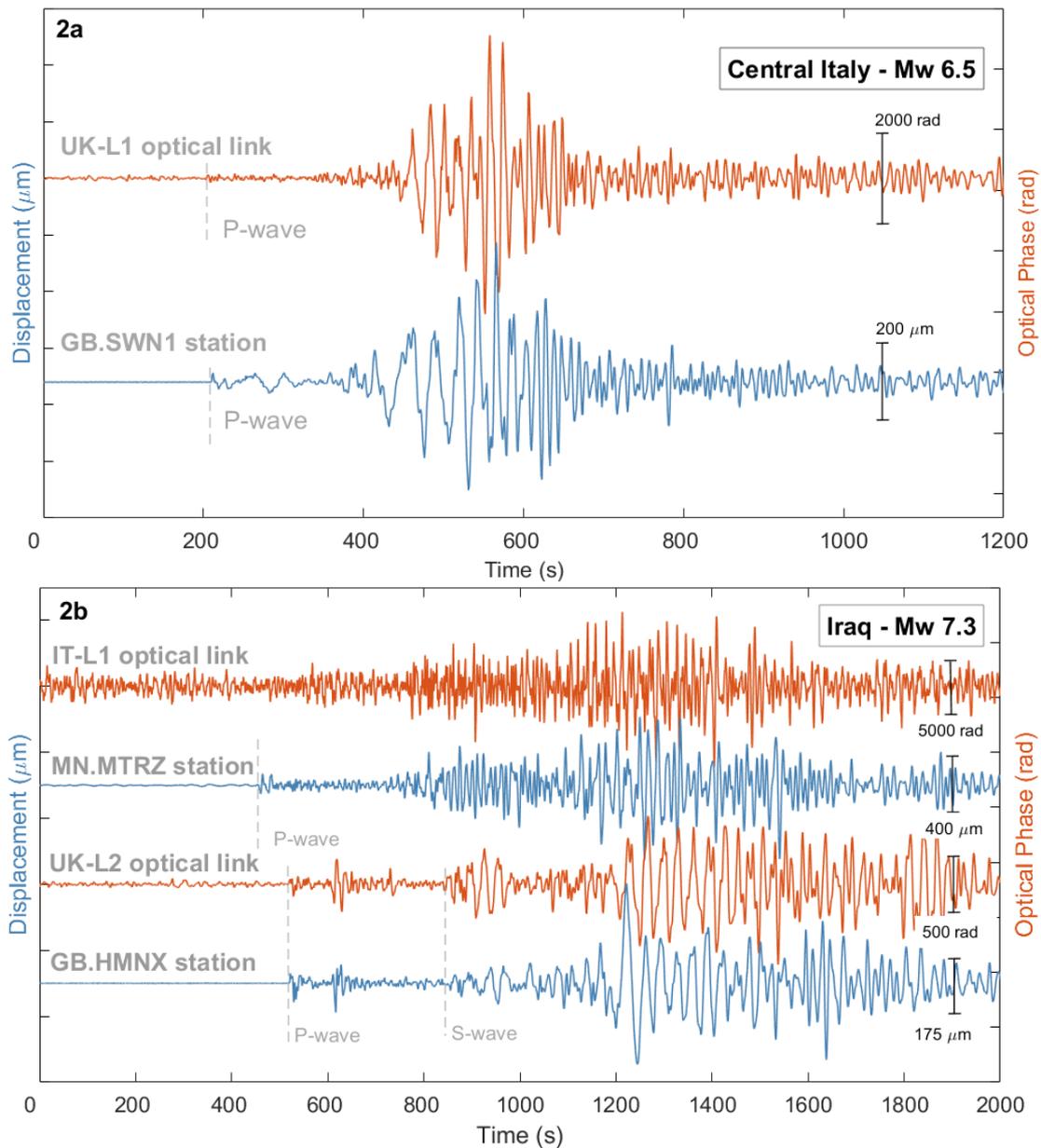

**Figure 2 | Tele-seismic events on terrestrial optical links. a**, Comparison between the seismically-induced optical phase changes detected on the UK-L1 link and the signal from a seismometer in Swindon (GB.SWN1) for the Central Italy earthquake on 30th October 2016; **b**, Comparison between the phase changes detected on the IT-L1 and UK-L2 link with the signals from seismometers in Monterenzio (MN.MTRZ) and Herstmonceux (GB.HMNX) for the Iraq border earthquake on 12th November 2017. The north-south component has been used for all seismic station data.

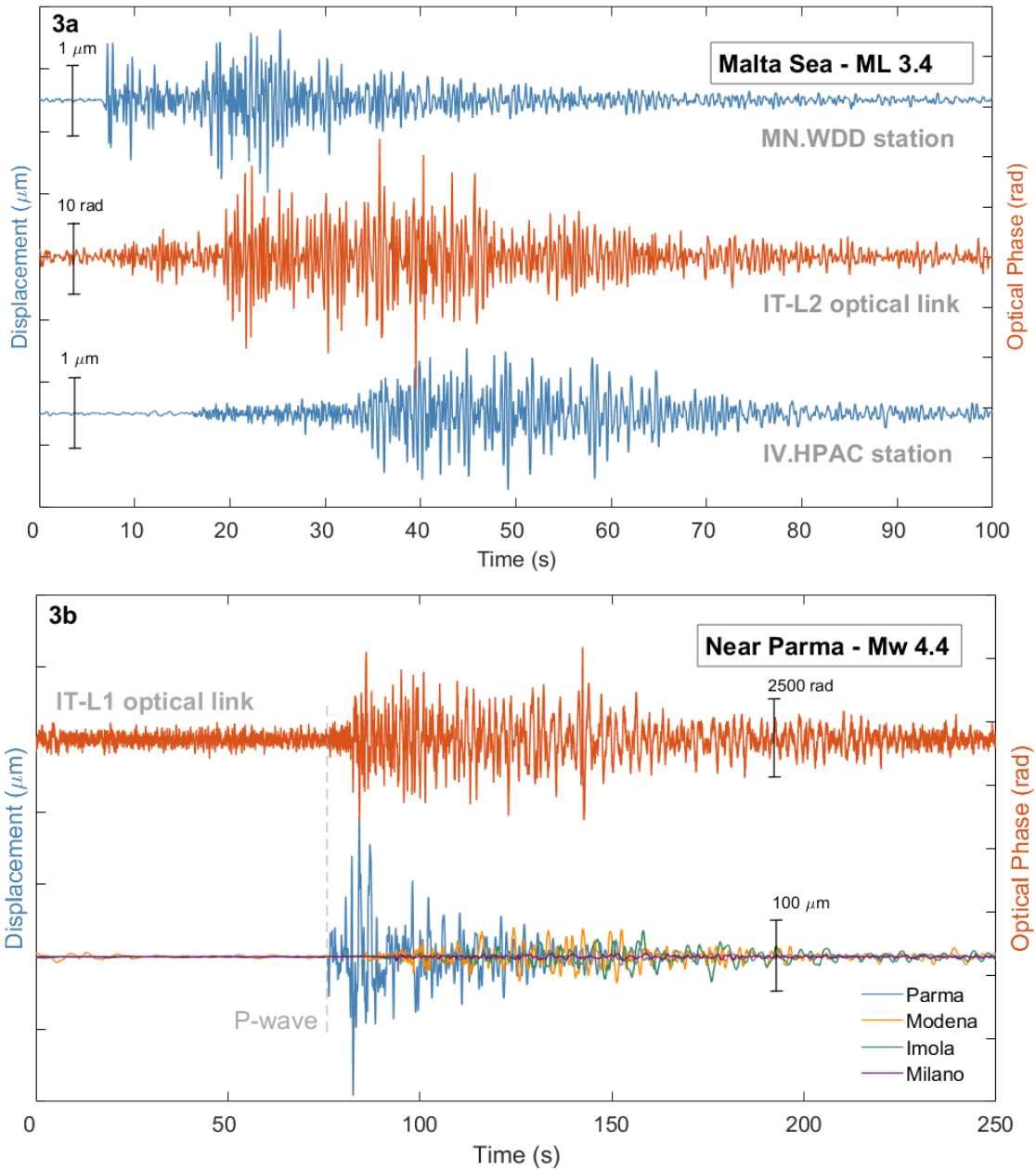

**Figure 3 | Small magnitude earthquakes on submarine and terrestrial link. a**, Seismic wave detected on the submarine IT-L2 link for the Malta Sea earthquake on 2nd September 2017 compared with signals from seismometers located within a few km of each end of the link. High pass filtering at 1.5 Hz has been applied to the optical signal to suppress a strong environmentally-induced 1 Hz component that was present on the optical signal. The same filter has been applied on the signals from seismic station. **b**, Comparison between the phase changes detected on IT-L1 link and seismometers located close to four intermediate points along the link for the Parma earthquake on 19th November 2017. The north-south component has been used for all seismic station data.

As with seismometers, the detection sensitivity of a terrestrial optical fibre link in the frequency range of interest for earthquake detection (0.1 to 20 Hz) is primarily limited by surrounding man-made noise. Substantially lower background noise per unit length is expected over submarine optical links, enabling the detection of low magnitude earthquakes. To test this hypothesis, INRIM, in collaboration with the University of Malta, conducted the first metrology experiments with an ultra-stable laser source on a submarine link in September 2017 (IT-L2). During a 2-day measurement campaign on the 98 km-long submarine cable between Malta and Sicily a local magnitude (ML) 3.4 earthquake, with epicentre in the Malta Sea, was detected. The distance from the epicentre to the nearest point on the link was approximately 85 km. The measured optical phase perturbation is shown in Fig. 3a and is compared to the displacement recorded by seismometers located within a few km of each end of the fibre link (MN.WDD, Malta and IV.HPAC, Sicily). A delay of approximately 2 s is observed between the P-wave detected by the MN.WDD station and that on the link, consistent with the travel distance between seismometer and the Malta end of the fibre link at a speed of approximately 5 km/s, which is calculated from the delay observed between the MN.WDD and IV.HPAC seismograms. Both P and S waves can be clearly identified. The root mean square level of environmental noise of the IT-L2 submarine link was measured to be 8 and 5 times lower than the UK-L1 and UK-L2 links respectively in the frequency range 0.1 to 20 Hz. The environmentally-induced noise, for spatially uncorrelated perturbations, scales with the length $L$ of the fibre link as $L^{1/2}$ [23]. Assuming similar levels of environmental noise, a 5,000 km-long trans-Atlantic link would therefore be only 7 times noisier than the IT-L2 link. Moreover, a quieter environment can be expected for cables resting on the ocean floor[24], which can reach depths of 6,000 m, than that found at the shallow depths (200 m) of the busy Malta-Sicily channel. Submarine cables cross several seismically active areas, such as the North and Mid-Atlantic ridge (including the triple junction where the South American, North American, and African Plates meet) the Indian Ocean and Indonesia. Seismic monitoring of all these areas relies almost entirely on seismic stations on land. Earthquakes of magnitude lower than 4 are largely undetected as they are too weak by the time they reach seismometers on the nearest island or mainland. Such earthquakes can typically be detected only up to a few hundred km from the epicentre, and so would affect only a relatively small fraction of the typical length of submarine fibre links. A similar scenario, on a smaller scale, occurred for detection of the Mw 4.4 Parma (Italy) earthquake on the IT-L1 link in November 2017. Here, the epicentre was 25 km away from the nearest section of the 535 km-long fibre link. Figure 3b shows the seismic wave as detected on the link in comparison with signals from four seismometers located in proximity to intermediate points along the optical link. The time of arrival of the seismic wave can be identified and corresponds to the smaller distance between the epicentre and the fibre link, which also leads to the highest amplitude of the detected signal.

The point at which the seismic wave reaches a fibre link can be determined by transmitting the laser light in both directions on separate fibres (optical telecommunication links always consist of fibre pairs), as depicted in Fig 4a. By cross-correlating the seismic signals recorded at each end of the link with a high-speed phase sampler referenced to the Coordinated Universal Time (UTC), the difference in travel time can be measured and thus the corresponding distance along the fibre[25]. Using this technique, in a proof-of-concept experiment, we were able to identify the location of an environmental perturbation to within 6 km over 126 km of spooled fibre in the laboratory. Two fibre links, following different paths, allow for the location of the epicentre to be found, as illustrated in Fig. 4b. The exact route of each fibre, required to calculate the epicentre, is normally known to within 1 km.

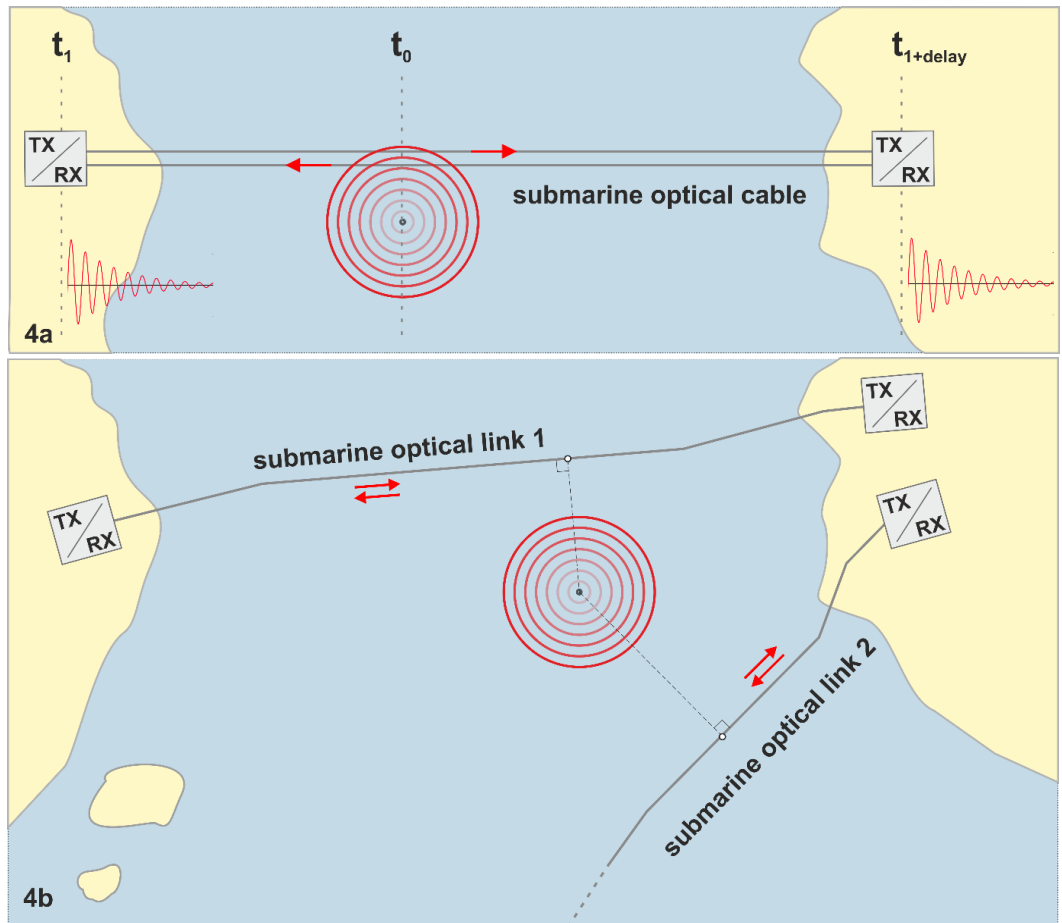

**Figure 4 | Localization techniques. a**, Because of the finite propagation speed of the light in the fibre (~$2 \times 10^8$ m/s) seismically-induced optical phase perturbations will reach the two ends of a bidirectional fibre link at different times. The location on the optical fibre link at which the seismic wave first reaches it can be determined by calculating the delay difference by cross-correlating the received signals. **b**, Localization of the epicentre using two bidirectional fibre links. Simple geometry allows the coordinates of the epicentre to be found from the location of the point of first contact of the seismic wave along the fibre. TX: ultra-stable laser injecting light into the fibre. RX: optical detection and phase comparison unit. This unit measures the optical phase difference between the light generated by the local TX laser and that transmitted through the fibre link by the remote TX laser.

The work presented here lays the foundation for the realization of a permanent global monitoring network that will enable real-time detection of a much larger number of submarine earthquakes than possible today, enabling new advances in our understanding of the underwater seismic activity of the Earth. We stress that our technique can be implemented on existing optical fibre infrastructure and requires only a single telecommunication channel whilst data traffic can be present on the other channels. These factors make it a cost effective solution for a permanent and low maintenance global monitoring network. The already extensive coverage of seas and oceans will continue to grow along with the demand for more connectivity and for new routes to be established for resilience. Even remote regions such as the Arctic are now also being crossed by submarine cables for the first time[26], with Antarctica being the only continent yet to be reached by a submarine cable. In future, submarine links could enable the early detection of tsunami-generating earthquakes by detecting them closer to the epicentre, gaining precious life-saving warning time. We anticipate that submarine fibre networks could also be used for applications beyond seismic monitoring, from marine mammal migration tracking[27] to sea noise pollution monitoring[27,28], a growing matter of concern world-wide for its impact on marine life.

**Acknowledgements** The UK side of this work was funded by the Department for Business, Energy and Industrial Strategy (BEIS) as part of the UK National Measurement System programme. The UK-L1 fibre link was funded by the UK Space Agency. INRIM acknowledges for funding the Italian Ministry of Education, University and Research (MIUR) through the Progetti Premiali 2014 and 2015 programs (LABMED and METGESP projects). We acknowledge funding by the Research, Innovation and Development Trust of the University of Malta. We are very grateful to Melita Limited and Enemalta plc for providing us with access to the submarine fibre links. We thank Prof. Massimo Inguscio, President of the Consiglio Nazionale delle Ricerche (CNR), for supporting and encouraging the Italy-Malta experiment. We thank Helen Margolis (NPL), Salvatore Micalizio (INRIM) and Prof. Pauline Galea (University of Malta) for fruitful discussions. The facilities of IRIS Data Services, and specifically the IRIS Data Management Center, were used for access to waveforms, related metadata, and/or derived products used in this study. IRIS Data Services are funded through the Seismological Facilities for the Advancement of Geoscience and EarthScope (SAGE) Proposal of the National Science Foundation under Cooperative Agreement EAR-1261681

**Author Contributions** G.M planned, designed and conducted the experiments on the UK links and prepared the first draft of the manuscript. R.L. and B.B. analysed the seismic data and provided seismology expertise. D.C. and F.L. planned and designed the experiments on IT-L1 and IT-L2. C.C., A.M., F.L and A.T. conducted the experiments on the IT-L1 link. C.C., D.C., A. X., and A.T. conducted the experiments on the IT-L2 link. L.W. developed the analytical tools. J.K. set up the UK-L2 link. S.R. provided underwater acoustic expertise. All authors contributed extensively to the discussion, interpretation of the data and manuscript preparation.

**Author Information** The authors declare no competing financial interests. Correspondence and requests for data should be addressed to G.M. (giuseppe.marra@npl.co.uk).

## METHODS SUMMARY

**The laser source** Narrow-linewidth continuous-wave telecommunication lasers (1530 to 1560 nm) phase locked to a hydrogen maser-referenced optical frequency comb were used in the measurement shown in Fig. 2a and to ultra-low-expansion (ULE) glass Fabry-Perot cavities[29] in all the other measurements shown. The fractional frequency stabilities of these lasers at 1 s averaging time are parts in $10^{13}$ and $10^{15}$ respectively. A fibre-laser phase-locked to a transportable ULE cavity was used for the measurements on the submarine link.

**Optical links** All the optical links used in the experiment consist of a pair of standard telecommunication optical fibres with a fully optical path between the transmitting and receiving end of the fibre links. Unidirectional or bidirectional Erbium-doped fibre amplifiers (EDFAs) were used every 70-100 km for optical signal regeneration. These are normally installed in data centres or fibre huts along the link. Digital data transfer signals were present on other channels for the UK-L1 and part of the IT-L1 links, whilst no additional signals were present on the other links. For UK-L1, UK-L2 and IT-L2 the laser light is injected in one of two fibres of the pair and returned on the other. For IT-L1 the laser light is retro-reflected in the same fibre; on this link, an acousto-optic modulator (AOM) at Medicina shifts the frequency of the incoming light before it is reflected. In this way the reflected signal can be distinguished from spurious localized reflections arising from interconnections along the link as well as distributed Rayleigh scattering.

**Optical phase detection** The light from the laser source is injected into the fibre link and returns to the laboratory after a round-trip. A portion of the injected light is frequency shifted (by a few tens of MHz) with an acousto-optic modulator (AOM) and combined with the returned light on a photodiode. A radio-frequency (RF) signal at the AOM frequency is generated at the output of the photodiode by interference of the two optical signals. The phase changes of the RF signal, which arise from accumulated optical phase changes of the light travelling in the fibre link, are measured with a digital phase meter with a sample rate of up to 500 samples/s.

**Perturbation localization test** The light from a ULE cavity-stabilized laser is injected, using an optical power splitter, in both directions in a 126 km-long loop, formed by two 50 km, one 1 km and one 25 km-long fibre spools, and an acousto-optic modulator (80 MHz). For each direction of travel, after a round trip the optical signal is combined with the light available at the output of the laser and a 80 MHz interferometric signal is obtained with a photodiode. This signal is down-converted to 100 kHz, sampled at 2.5 Msamples/s and its phase is extracted by I/Q demodulation. An environmental perturbation is created by tapping on the 1 km spool. By performing the cross-correlation of the two interferometric signals, the difference in travel time in clockwise and counter-clockwise direction from the perturbation location to the photodiodes is extracted. From this delay the location of the perturbation can be calculated by multiplying the delay by the speed of light in the fibre (~2 x $10^8$ m/s).

**Epicentre localization** A standard telecommunication fibre link consists of a fibre pair, one for each direction of propagation of the light. Unidirectional optical amplifiers are used along the link on each of the fibre pair. At each end of a fibre link, a ULE-cavity stabilized laser (TX) injects phase-stable light into one fibre of pair and a receiver unit (RX) combines the incoming light available on the other fibre (transmitted by the remote TX laser) with that from the local TX laser. An interferometric signal is produced on a photodiode at both ends of the fibre and acquired at a fast sampling rate with a UTC-referenced analog-to-digital converter. The location of the perturbation along the fibre is extracted using the method described in the perturbation localization test. Synchronization of the converter's time base to UTC at the 100 ns level can be achieved with GPS-disciplined oscillators at either end of the fibre. Alternatively a single GPS-disciplined oscillator can be used at one end of the fibre and UTC time can be transferred to the other end using time transfer techniques[30]. By using two fibre links following different paths (Figure 4b) the location of the epicentre can be found by intersecting the normal to each fibre at the location of the first point of contact of the seismic wave with the fibre.

**Data availability.** All relevant data are available from the authors upon reasonable Request.